\def\beg{\begin{equation}}
\def\eeq{\end{equation}}
\documentstyle[12pt]{article}
\begin{document}
\begin{center}
{\Large{\bf New Zero-Resistance States in Heterostructures: 
Anderson-Brinkman-Phillips Charge-Density Waves and New States}}
\vskip0.35cm
{\bf Keshav N. Shrivastava}
\vskip0.25cm
{\it School of Physics, University of Hyderabad,\\
Hyderabad  500046, India}
\end{center}

Recently observed zero-resistance states are interpreted by 
Anderson-Brinkman and Phillips as charge density waves. We 
point out the existence of charge-density waves,
 superconducting states and new 
states including a state of zero charge and finite spin.
\vskip1.0cm
Corresponding author: keshav@mailaps.org\\
Fax: +91-40-2301 0145.Phone: 2301 0811.
\vskip1.0cm

\noindent {\bf 1.~ Introduction}

     Recently zero-resistivity has been observed[1,2] 
at small magnetic fields when the heterostructures are 
irradiated by microwaves. Phillips has suggested that 
the observation may be explained by sliding charge 
density waves which can give very low resistivity in 
the presence of a small magnetic field. Anderson and 
Brinkman have considered two 
possible mechanisms. According to one of the mechanisms, 
there are both positive as well as negative contributions 
to the resistivity so that the resultant value comes 
near zero. The negative resistance may imply charge-density 
fluctuations which may be observable in the power spectrum. 
In the past whenever a near zero resistance was found, it was 
assigned to charge-density waves and in a few cases, 
superconductivity was identified by using Meissner effect. 
When both the experiments, i.e., the zero-resistivity and 
negative susceptibility were successful, it was thought to 
be superconductivity with spin-singlet states. In recent years, 
a complex gap of a superconductor has been discussed. However, 
in the case of $^3He$, singlets, triplets with $\pm$ 1 
component, and triplets with 0 as well as $\pm$ 1 components 
have been found to phase separate.   

    Anderson-Brinkman[3] as well as Phillips[4] agree that 
there are some kind of charge density waves. We have been 
working on such 
problems for some time so that we pray that our interpretation 
may
 also be considered. We are not against the charge-density waves 
which is one of the components in our solution. Our solution 
follows.

\noindent{\bf 2.~~Description}

     When we try to calculate the effective magnetic moment we 
find several solutions[5]. One of the solutions has zero g-value 
so that 
the frequency of this mode is zero. This solution is therefore 
similar to the charge-density waves. When spin is reversed, the 
zero energy solution changes to a finite energy solution. If we 
can have one particle with spin up and zero energy and the other 
particle with spin down and finite energy, we can have a new type 
of superconductivity so that the resistivity is zero due to zero 
orbital angular momentum and zero spin. The orbital angular 
momentum of each of the particles is zero, ${\it l}=0$, and 
spin for one is $down$ while for the other is $up$ so that 
there is a zero momentum, spin-singlet type state which is 
superconducting. Thus for ${\it l}=0$, only these two solutions 
are found, namely, the charge density waves or the 
superconductivity. These states are distinguished by the 
sign of susceptibility. The charge-density waves are paramagnetic 
whereras superconductivity is diamagnetic.

     Let us now imagine that {\it l} becomes finite. Here the 
contribution of Anderson model becomes very important. It may
 mean that going from {\it l}=0 to {\it l}=1 is an Anderson type 
transition. Once {\it l}=1, we have an interesting situation in 
which one spin configuration  such as -1/2 gives an effective 
charge of 1/3 and +1/2 gives a charge of 2/3. In fact, varying 
the values of {\it l} reproduces ``Stormer's" values[7] exactly.
 It is a wonder that this theory gives all of the fractional 
charges exactly as found in the experimental data on the 
quantum Hall effect. For very large values of the orbital angular 
momentum, even the 1/2 filled Landau level is produced and there 
is a Bose-Einstein condensation at this point. The width of the 
plateau is determined by the ordinary electron-phonon type 
interaction which can be reduced to an activation energy type 
process.

     In this way, we have many solutions, one of which is a 
charge density, the other is superconductivity and there are 
others with effective fractional charges. This point of view is 
slightly different from Anderson-Brinkman-Phillips interpretation, 
yet it has all of the features of the experimental data. We have
 a particle[6] with zero charge and spin 1/2 so that its frequency 
is also zero which is a characteristic of the charge-density wave,
 in agreement with Anderson, Brinkman and Phillips.
 
\noindent{\bf3.~~ Conclusions}.

     Some new states occur. These states are superconducting 
but as the magnetic field is varied, the spin flips and the 
state is destroyed and then another state is created.  One of 
the solutions has a charge density wave with zero frequency but
 the resistivity results are dominated by non-charge density 
waves. Therefore, the results of 
refs.1 and 2 have new results[5].

\vskip1.25cm

\noindent{\bf4.~~References}
\begin{enumerate}
\item R. G. Mani, et al, Nature {\bf419}, 646(2002)
\item M. A. Zudov, et al,  Phys. Rev. Lett. {\bf 90}, 046807(2003).
\item P. W. Anderson and W. F. Brinkman, cond-mat/0302129.
\item J. C. Phillips, cond-mat/0212416.
\item K.N. Shrivastava, Introduction to quantum Hall effect,\\ 
      Nova Science Pub. Inc., N. Y. (2002).
\item K. N. Shrivastava, cond-mat/0212552.
\item H. L. Stormer, Rev. Mod. Phys. {\bf 71}, 875 (1999).
\end{enumerate}
\vskip0.1cm

Note: Ref.5 is available from:\\
 Nova Science Publishers, Inc.,\\
400 Oser Avenue, Suite 1600,\\
 Hauppauge, N. Y.. 11788-3619,\\
Tel.(631)-231-7269, Fax: (631)-231-8175,\\
 ISBN 1-59033-419-1 US$\$69$.\\
E-mail: novascience@Earthlink.net

\end{document}